\documentclass{kapproc} 

\def\ls{{_<\atop^{\sim}}}
\def\gs{{_>\atop^{\sim}}}
\def\cgs{ ${\rm erg~cm}^{-2}~{\rm s}^{-1}$ } 
\def\aap{A\&A}
\def\apj{ApJ}
\def\aj{AJ}

\usepackage{procps} 

\usepackage[dvips]{graphicx}

\upperandlowercase

\normallatexbib 

\begin{document}

\articletitle[The HELLAS2XMM survey] {The HELLAS2XMM 1dF survey: on
the nature of high X-ray to optical flux ratio sources
}

\author{F. Fiore, M. Brusa, F. Cocchia, A. Baldi, N. Carangelo,
P. Ciliegi, A. Comastri, F. La Franca, R. Maiolino, G. Matt,
S. Molendi, M. Mignoli, G.C. Perola, P. Severgnini, C. Vignali}




\anxx{Fiore\, F.}

\begin{abstract}
We present results from the photometric and spectroscopic
identification of 122 X-ray sources recently discovered by XMM-Newton
in the 2-10 keV band (the HELLAS2XMM 1dF sample).  One of the most
interesting results (which is found also in deeper sourveys) is that
$\sim20\%$ of the sources have an X-ray to optical flux ratio (X/O)
ten times or more higher than that of optically selected AGN.  Unlike
the faint sources found in the ultra-deep Chandra and XMM-Newton
surveys, which reach X-ray (and optical) fluxes $\gs10$ times lower
than in the HELLAS2XMM sample, many of the extreme X/O sources in our
sample have R$\ls25$ and are therefore accessible to optical
spectroscopy.  We report the identification of 13 sources with extreme
X/O values. While four of these sources are broad line QSO, eight of
them are narrow line QSO, seemingly the extension to very high
luminosity of the type 2 Seyfert galaxies.

\end{abstract}


\section{Introduction}

Hard X-ray surveys are the most direct probe of supermassive black
hole (SMBH) accretion activity, which is recorded in the Cosmic X-ray
Background (CXB), in wide ranges of SMBH masses, down to $\sim
10^6-10^7\,M_{\odot}$, and bolometric luminosities, down to $L\sim
10^{43}$ erg/s. X-ray surveys can therefore be used to: study the
evolution of the accreting sources; measure the SMBH mass density;
constrain models for the CXB \cite{setti89,coma95}, and models
for the formation and evolution of the structure in the universe
\cite{haehnelt03,menci03}.  These studies have so far confirmed, at
least qualitatively, the predictions of standard AGN synthesis models
for the CXB: the 2-10 keV CXB is mostly made by the superposition of
obscured and unobscured AGNs (\cite{hasi03,fiore03a} and references
therein).  Quantitatively, though, rather surprising results are
emerging: a rather narrow peak in the range z=0.7-1 is present in the
redshift distributions from ultra-deep Chandra and XMM-Newton
pencil-beam surveys, in contrast to the broader maximum observed in
previous shallower soft X-ray surveys made by ROSAT, and predicted by
the above mentioned synthesis models.  However, the optical
identification of the faint sources in these ultra-deep surveys is
rather incomplete, especially for the sources with very faint optical
counterparts, i.e. sources with high X-ray to optical flux ratio
(X/O). Indeed, the optical magnitude of $\approx20\%$ of the sources,
those having the higher X/O, is R$\gs26-27$, not amenable at present
to optical spectroscopy.  This limitation leads to a strong bias in
ultra-deep Chandra and XMM-Newton surveys against AGN highly obscured
in the optical, i.e. against type 2 QSOs, and in fact, only 10 type 2
QSOs have been identified in the CDFN and CDFS samples
\cite{cowie03,hasi03}.  
To help overcoming this problem, we are pursuing a large area,
medium-deep surveys, the HELLAS2XMM serendipitous survey, which, using
XMM-Newton archival observations \cite{baldi02} has the goal to cover
$\sim4$ deg$^2$ at a 2-10 keV flux limit of a few$\times10^{-14}$
\cgs. At this flux limit several sources with X/O$\gs10$ have optical
magnitudes R=24-25, bright enough for reliable spectroscopic redshifts
to be obtained with 10m class telescopes.

\section {The HELLAS2XMM 1dF sample}

We have obtained, so far, optical photometric and spectroscopic
follow-up of 122 sources in five XMM-Newton fields, covering a total
of $\sim0.9$ deg$^2$ (the HELLAS2XMM `1dF' sample), down to a flux
limit of F$_{2-10keV}\sim10^{-14}$ erg cm$^{-2}$ s$^{-1}$.  We found
optical counterparts brighter than R$\sim25$ within $\sim6''$ from the
X-ray position in 116 cases and obtained optical spectroscopic
redshifts and classification for 94 of these sources
\cite{fiore03}. The source breakdown includes: 61 broad line QSO and
Seyfert 1 galaxies, and 33 {\em optically obscured AGN}, i.e.  AGN
whose nuclear optical emission, is totally or strongly reduced by dust
and gas in the nuclear region and/or in the host galaxy (thus
including objects with optical spectra typical of type 2 AGNs,
emission line galaxies and early type galaxies, but with X-ray
luminosity $\gs10^{42}$ erg s$^{-1}$).
We have combined the HELLAS2XMM 1dF sample with other deeper hard
X-ray samples including the CDFN \cite{barger02}, Lockman Hole
\cite{main02,baldi02}, and SSA13 \cite{barger01} samples, to collect a
``combined'' sample of 317 hard X-ray selected sources, 221 (70\%) of
them identified with an optical counterpart whose redshift is
available.  The flux of the sources in the combined sample spans in
the range $10^{-15}-4\times10^{-13}$ \cgs and the source breakdown
includes 113 broad line AGN and 108 optically obscured AGN.

\begin{figure}[h]
\includegraphics[height=7cm,angle=-90]{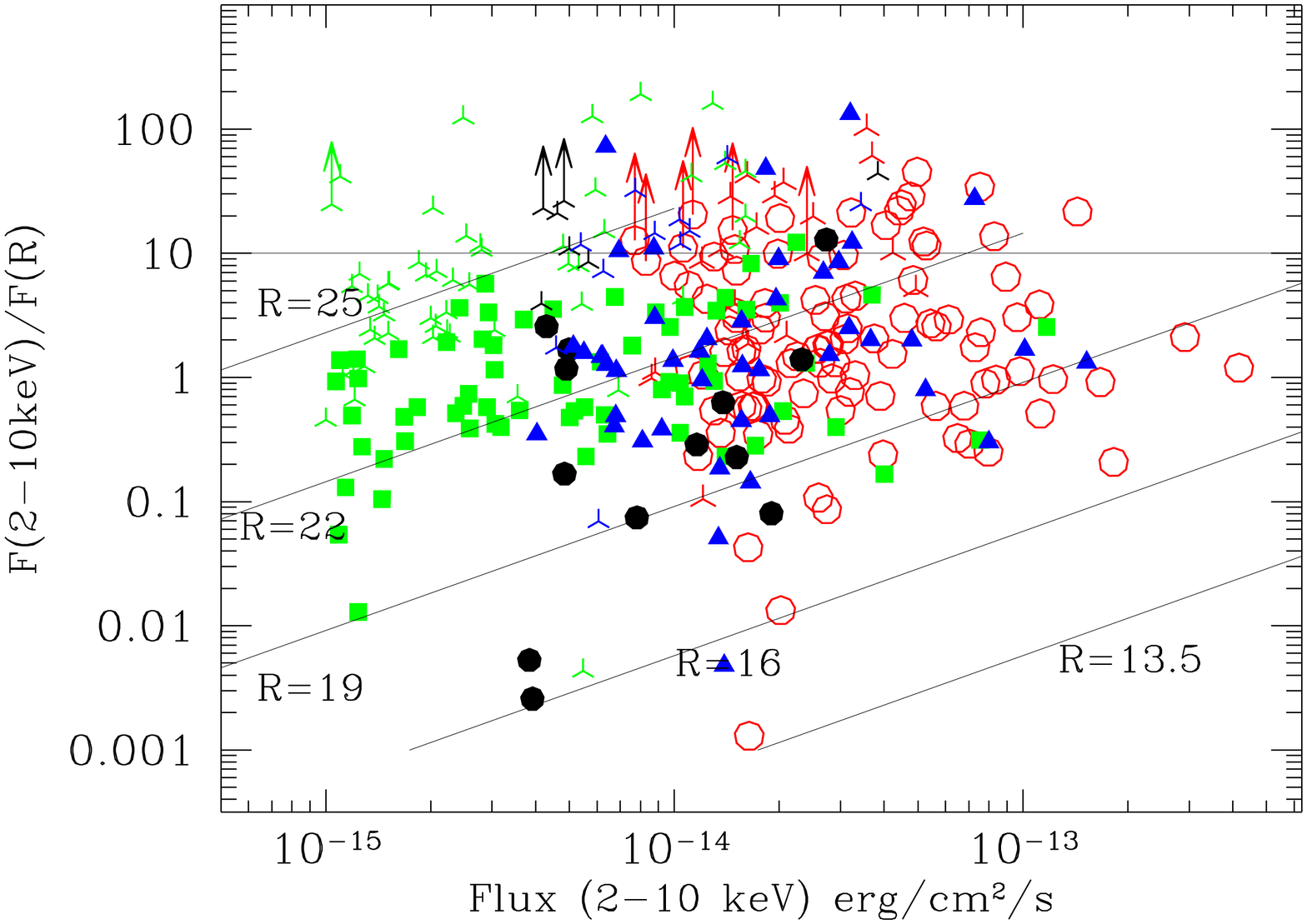}
\vskip -5.7truecm
\narrowcaption{The X-ray (2-10 keV) to optical (R band) flux ratio X/O as a
function of the X-ray flux for the combined sample (HELLAS2XMM = open
circles; CDFN = filled squares; LH = filled triangles; SSA13 = filled
circles, skeleton triangles are sources without a measured redshift).
Solid lines mark loci of constant R band magnitude.}
\label{xos}
\vskip -0.5truecm
\end{figure}

\begin{figure}[h]
\includegraphics[height=12cm,angle=-90]{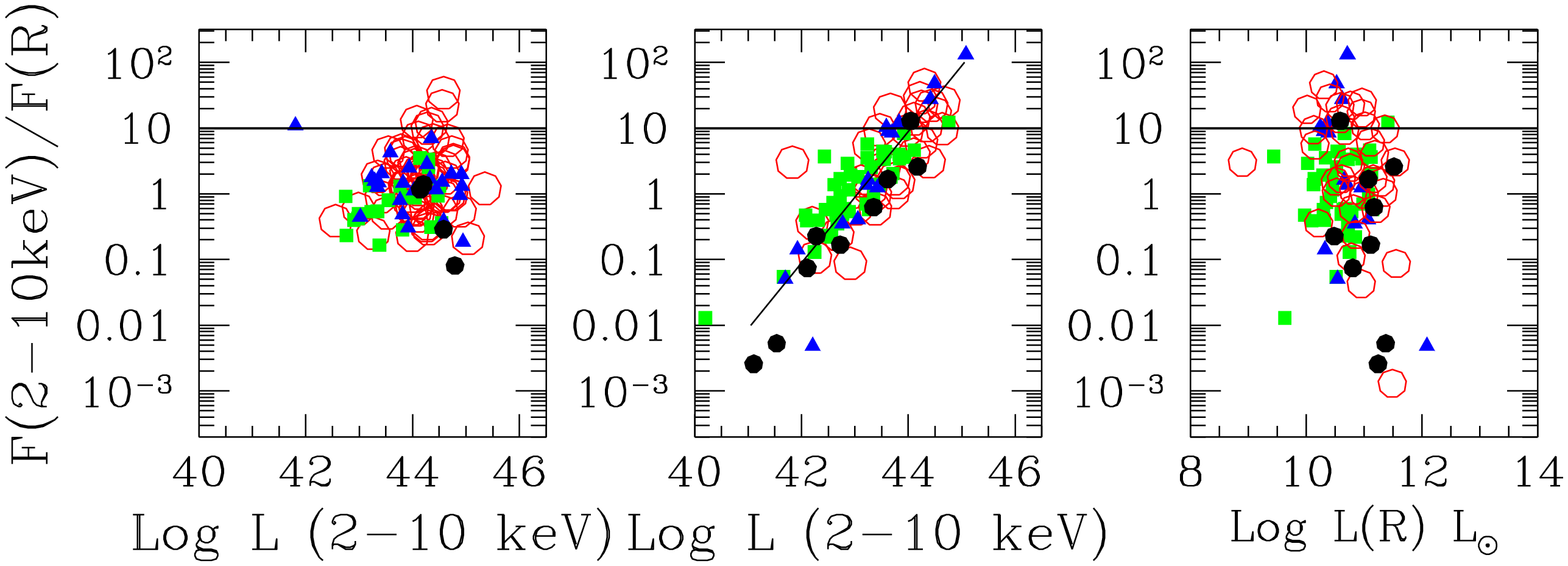}
\vskip -0.5truecm
\caption{The X-ray to optical flux ratio X/O versus the X-ray
luminosity for type 1 AGN (left panel), and non type 1 AGN and
galaxies (central panel); X/O versus the optical luminosity for non
type 1 AGN and galaxies (right panel).  Symbols as in
Fig. \ref{xos}. The orizontal lines mark the level of X/O=10,
$\sim20\%$ of the sources in the combined sample have X/O higher than
this value. The diagonal line in the right panel is the best
log(X/O)--log(L$_{2-10keV}$) linear regression.}
\label{xolx}
\end{figure}

Fig. \ref{xos} shows the X-ray (2-10 keV) to optical (R band) flux
ratio (X/O) as a function of the hard X-ray flux for the combined sample.  
About 20\% of the sources have
X/O$\gs10$, i.e ten times or more higher than the X/O typical of
optically selected AGN.  At the flux limit of the HELLAS2XMM 1dF
sample several sources with X/O$\gs10$ have optical magnitudes
R=24-25, bright enough to obtain reliable spectroscopic redshifts.
Indeed, we were able to obtain
spectroscopic redshifts and classification of 13 out of the 28
HELLAS2XMM 1dF sources with X/O$>10$; {\em 8 of them are type 2 QSO at
z=0.7-1.8}, to be compared with the total of 10 type 2 QSOs identified
in the CDFN \cite{cowie03} and CDFS \cite{hasi03}.  Fig. \ref{xolx}
show the X-ray to optical flux ratio as a function of the X-ray
luminosity for broad line AGN (left panel) and non broad line AGN and
galaxies (central panel).  While the X/O of the broad line AGNs is not
correlated with the luminosity, a striking correlation between
log(X/O) and log(L$_{2-10keV}$) is present for the obscured AGN:
higher X-ray luminosity, optically obscured AGN tend to have higher
X/O. A similar correlation is obtained computing the ratio between the
X-ray and optical luminosities, instead of fluxes (because the
differences in the K corrections for the X-ray and optical fluxes are
small in comparison to the large spread in X/O).  All objects plotted
in the right panel of Fig. \ref{xolx} do not show broad emission
lines, i.e. the nuclear optical-UV light is completely blocked, or
strongly reduced in these objects, unlike the X-ray light. Indeed, the
optical R band light of these objects is dominated by the host galaxy
and, therefore, {\em X/O is roughly a ratio between the nuclear X-ray
flux and the host galaxy starlight flux}.  The right panel of Figure
\ref{xolx} helps to understand the origin of the correlation between
X/O and L$_{2-10keV}$. While the X-ray luminosity of the optically
obscured AGNs spans about 4 decades, the host galaxy R band luminosity
is distributed over less than one decade.  The ratio between the two
luminosities (and hence the ratio between the two fluxes, see above)
results, therefore, strongly correlated with the X-ray luminosity.

\section{Summary}

We have obtained spectroscopic redshifts and classification of 13 out
of the 28 HELLAS2XMM 1dF sources with X/O$\gs10$: the majority of
these sources (8) are type 2 QSOs at z=0.7-1.8, a fraction of type 2
QSOs much higher than at lower X/O values.  We find a strong
correlation between X/O and the X-ray luminosity of optically obscured
AGN, X/O=10 corresponding to an (average) 2-10 keV luminosity of
$10^{44}$ erg s$^{-1}$.  Sources of this luminosity and flux $\approx
10^{-15}$ \cgs, reachable in Chandra and XMM-Newton ultra-deep
surveys, would be at z$\sim3$. Although only 20\% of the X-ray sources
have such high X/O, they may carry the largest fraction of accretion
power from that shell of Universe.  Intriguingly, Mignoli et al. (2003
in preparation) find a strong correlation between the R-K color and
the X/O ratio for a selected sample of 10 high X/O HELLAS2XMM 1dF sources,
all of them  
having R-K$\gs5$, i.e. they are all Extremely Red Objects.


\begin{chapthebibliography}{1}
\bibitem {setti89} Setti, G., \& Woltjer, L. 1989, \aap, 224, L21
\bibitem{coma95} Comastri, A., Setti, G., Zamorani, G., \& Hasinger, G. 1995, 
\aap, 296, 1   
\bibitem{haehnelt03} Haehnelt, M. Carnegie Observatories 
Astrophysics Series, Vol. 1: Coevolution of Black Holes and Galaxies, 
ed. L. C. Ho (Cambridge Univ. Press), 2003, astro-ph/0307378
\bibitem{menci03} Menci, N. et al. 2003, \apj , 587, L63
\bibitem{hasi03} Hasinger, G. 2003, proceedings of the Conference: 
The Emergence of Cosmic Structure, Maryland, Stephen S. Holt and 
Chris Reynolds (eds), astro-ph/0302574 
\bibitem{fiore03a} Fiore, F. 2003, proceedings of the symposium
"The Restless High-Energy Universe", E.P.J. van den Heuvel, 
J.J.M. in 't Zand, and R.A.M.J. Wijers Eds, astro-ph/0309355
\bibitem{cowie03} Cowie L., Barger A., Bautz, M.W., Brandt, W.N., \& Garnire,
G.P. 2003, \apj, 584, L57
\bibitem{fiore03} Fiore, F. Brusa, M, Cocchia, F. et al. 2003, A\&A
in press, astro-ph/0306556
\bibitem{baldi02} Baldi, A., Molendi, S., Comastri, A., Fiore, F., Matt,
G., \& Vignali, C. 2002, \apj, 564, 190
\bibitem{barger02} Barger A., et al. 2002, \aj, 124, 1839
\bibitem{barger01} Barger, A., Cowie, L., Mushotzky, R.F., \&
Richards, E.A. 2001, \aj, 121, 662 
\bibitem{main02} Mainieri, V. et al. 2002, \aap, 393, 425 
\end{chapthebibliography}
\end{document}